\documentclass[twocolumn,showpacs,prl]{revtex4}

\usepackage[dvips]{graphicx}% Include figure files
\usepackage{dcolumn}% Align table columns on decimal point
\usepackage{bm}% bold math

\begin{document}

\title{Electronic Transport Properties of the Ising Quantum Hall Ferromagnet
 in a Si Quantum Well}

\author{Kiyohiko Toyama,$^1$ Takahisa Nishioka,$^1$ Kentarou Sawano,$^2$ Yasuhiro Shiraki,$^2$ and Tohru Okamoto$^1$}
\affiliation{$^1$Department of Physics, University of Tokyo, 7-3-1, Hongo, Bunkyo-ku, Tokyo 113-0033, Japan\\
$^2$Research Center for Silicon Nano-Science, Advanced Research Laboratories, Musashi Institute of Technology, 8-15-1 Todoroki, Setagaya-ku, Tokyo 158-0082, Japan}

\date{26 February 2008}

\begin{abstract}
Magnetotransport properties are investigated for a high mobility Si two dimensional electron systems in the vicinity of a Landau level crossing point.
At low temperatures, the resistance peak having a strong anisotropy shows large hysteresis which is attributed to Ising quantum Hall ferromagnetism.
The peak is split into two peaks in the paramagnetic regime.
A mean field calculation for the peak positions indicates that electron scattering is strong when the pseudospin is partially polarized.
We also study the current-voltage characteristics which exhibit a wide voltage plateau.
\end{abstract}
\pacs{73.43.Nq, 73.40.Lq, 73.43.Qt, 75.60.-d}

\maketitle

In the presence of a perpendicular magnetic-field $B_\perp$, the single-particle energy spectrum of a two-dimensional electron system (2DES) is quantized into Landau levels (LLs) and the integer quantum Hall (QH) effect occurs at low temperatures.
The spin degree of freedom, as well as the subband degree of freedom in wide quantum wells, provides us an opportunity to study quantum phase transitions induced by crossing LLs with different orbital and spin (or subband) indices at the Fermi level \cite{Koch1993,Daneshvar1997,Jungwirth1998,Poortere2000,Jungwirth2000,Zeitler2001,Muraki2001,Jungwirth2001,Jaroszynski2002,Poortere2003,Chokomakoua2004,Lai2006}.
The two LLs that are nearly degenerate in energy can be relabeled in terms of pseudospin \cite{Jungwirth1998,Jungwirth2000,MacDonald1990}.
Electron-electron interaction stabilizes the ferromagnetic alignment of pseudospin and its dependence on the orbital index $n$ leads to easy-axis (Ising) or easy-plane ($XY$) anisotropy.
A useful method for LL crossing is to tilt the magnetic field since the ratio of the cyclotron energy $\hbar \omega_c=\hbar e B_\perp/m^\ast$ to the Zeeman energy gap $|g^\ast| \mu_B B_{\rm tot}$ changes with the tilt angle $\theta$.
Here $\omega_c$ is the cyclotron frequency, $m^\ast$ is the effective mass, $g^\ast$ is the effective $g$-factor, and $B_{\rm tot}$ is the total strength of the magnetic field.
For idealized single-layer QH systems in a tilted magnetic filed, the ferromagnetic ground state is predicted to have Ising anisotropy \cite{Jungwirth1998,Jungwirth2000}.
Resistance spikes and hysteresis have been observed for several 2DESs and discussed 
in association with domain wall resistance in Ising QH ferromagnets \cite{Poortere2000,Jaroszynski2002,Poortere2003,Chokomakoua2004},
while the data were obtained by sweeping the magnetic field at fixed angles.

In this Letter, we report systematic magnetotransport measurements on a very high mobility Si 2DES in the vicinity of the coincidence of the spin-up and spin-down LLs with $n=0$ and 1, respectively.
In order to sweep the pseudospin effective field, which is proportional to $\hbar \omega_c - g^\ast \mu_B B_{\rm tot}$, the data are obtained by continuously changing $\theta$ at a fixed magnetic field.
The resistance peak at the coincidence exhibits a strong anisotropy with respect to the angle $\varphi$ between the in-plane magnetic field and direction of the electric current.
The anisotropy factor reaches up to 50 at 50~mK.
Large hysteresis, which demonstrates the Ising QH ferromagnetism, is observed at low temperatures while it disappears as $T$ increases.
In the paramagnetic regime ($T \gtrsim 0.5$~K), the resistance peak is split into two peaks.
Using the mean field approach for the $T$-dependent peak positions, we deduce that strong electron scattering occurs when the pseudospin is partially polarized.
We also study the current-voltage characteristics.
A wide voltage plateau observed at 0.37~K is explained in terms of the breakdown of the uniform current distribution.

The sample used is a Si/SiGe heterostructure with a 20-nm-thick strained Si channel sandwiched between relaxed $\mathrm{Si}_{0.8}\mathrm{Ge}_{0.2}$ layers \cite{Yutani1996}.
The electrons are provided by a Sb-$\delta$-doped layer 20~nm above the channel.
Recent electron spin resonance measurements demonstrate that the spin-orbit interactions and the electron-nuclear spin (hyperfine) coupling are very small in the present system \cite{Matsunami2006}.
The 2D electron concentration $N_s$ was adjusted to $2.4\times 10^{15}~\mathrm{m^{-2}}$  with a mobility $\mu \approx 60~\mathrm{m^2/V~s}$ by illumination with a red light-emitting diode in different cooling runs.
The longitudinal resistivity $\rho_{xx}$ was measured for a current channel of total length 1.8~mm and width $w=0.2$~mm with voltage probes separated by $l=0.8$~mm as illustrated in Fig.~1(a).
To avoid the heating effect, a small ac current of 1~nA was used in a standard low-frequency (11.4~Hz) lock-in technique except for $I$-$V$ characteristics measurements.
The tilt angle $\theta$ of the sample with respect to the magnetic field ($B_\perp=B_{\rm tot} \cos \theta$) was controlled using a rotatory stage in a dilution refrigerator or in a pumped ${}^3$He refrigerator.

\begin{figure}[t!]
\includegraphics[width=8cm]{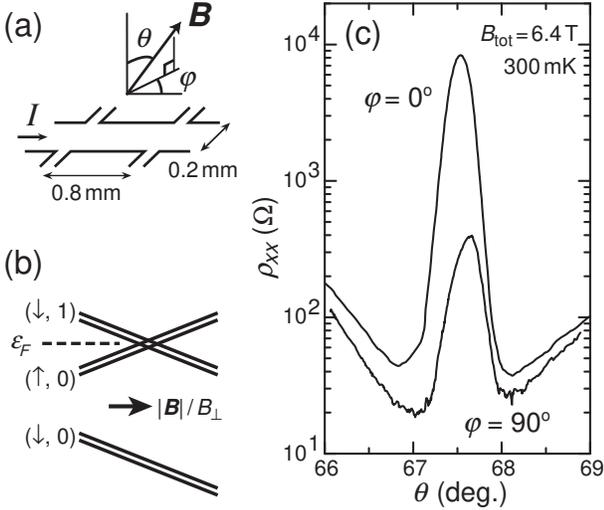}
\caption{
(a) Schematic view of the central part of the sample located in a tilted magnetic field. (b) Landau level configuration in a tilted magnetic field. (c) The longitudinal resistivity as a function of $\theta$ for $\varphi=0^\circ$ and $90^\circ$ at $T=300$~mK.
The magnetic field of 6.4~T is chosen so that $\nu$ equals to 4.0 at the resistance peak.
}
\end{figure}

At the coincidence of the spin-up LL with $n=0$ [LL$(\uparrow, 0)$] and the spin-down LL with $n=1$ [LL$(\downarrow, 1)$], the Fermi level $\varepsilon_F$ lies at these LLs when the Landau level filling factor $\nu$ is in the range between 2 and 6 [see Fig.~1(b)].
Note that we have a two-fold valley degeneracy for a 2DES formed on the Si(001) surface and the spin-down state being antiparallel to the magnetic field has lower Zeeman energy than the spin-up state for conduction electrons in Si with $g^\ast>0$.
The coincidence angle $\theta_c$, at which resistance anomaly was observed, exhibits a small $\nu$-dependence corresponding to a linear relationship
$B_\perp ({\rm T}) = 0.455 B_{\rm tot} ({\rm T}) - 0.44$ for $5~{\rm T} < B_{\rm tot} < 9~{\rm T}$.
The ratio $B_\perp/B_{\rm tot}$ is significantly enhanced from 0.19 for noninteracting electrons with $m^\ast=0.19 m_e$ and $g^\ast=2.00$ due to many body effects \cite{Okamoto1999,Pudalov2002,Okamoto2004}.
Here $m_e$ is the free electron mass.
The resistance peak at $\theta_c$ was found to be especially high around $\nu=4$.
In Fig.~1(c), $\theta$-dependence of $\rho_{xx}$ at $T= 300$~mK and $\nu=4.0$ is shown for different angles $\varphi$ of the in-plane magnetic field with respect to the current channel.
The peak height for $\varphi =0^\circ$ is much larger than that for $\varphi =90^\circ$ \cite{Crystallographic}.
Similar resistance peak and anisotropy in a Si/SiGe heterostructure have been reported by Zeitler {\it et al.} in their study in very high magnetic fields above 20~T \cite{Zeitler2001}.
The observation at a moderate magnetic field $B_{\rm tot} =6.4$~T in the present work may be attributed to the very high mobility of the sample.
The peak height at $\nu=4.0$ for $\varphi =0^\circ$ was found to be reduced down to 2~k$\Omega$ when the measurement was performed at $N_s = 1.4\times 10^{15}~\mathrm{m^{-2}}$  with $\mu = 34~\mathrm{m^2/V~s}$.
The origin of the anisotropic magnetoresistance observed in Ref.~\cite{Zeitler2001} has been discussed by Chalker {\it et al.} in terms of domain formation induced by fluctuations of the pseudospin field arising from isotropic surface roughness \cite{Chalker2002}.
However, their model does not seem to explain hysteresis behavior observed in the present work.
Furthermore, the anisotropy factor $\rho_{xx}(\varphi=0^\circ)/\rho_{xx}(\varphi=90^\circ)$ at the peak was found to reach up to 50 at 50~mK in contrast to the calculated values less than 10 \cite{Chalker2002}.
We speculate that Ising QH ferromagnetism stabilizes the anisotropic domain structure while the initial anisotropy may be given by the randomness of the pseudospin field.

The hysteresis behavior is observed at lower temperatures.
It is clearer for $\varphi=90^\circ$ while it appears also for $\varphi=0^\circ$.
In Fig.~2, the temperature evolution of the hysteresis for $\varphi=90^\circ$ is shown.
\begin{figure}[t!]
\includegraphics[width=8cm]{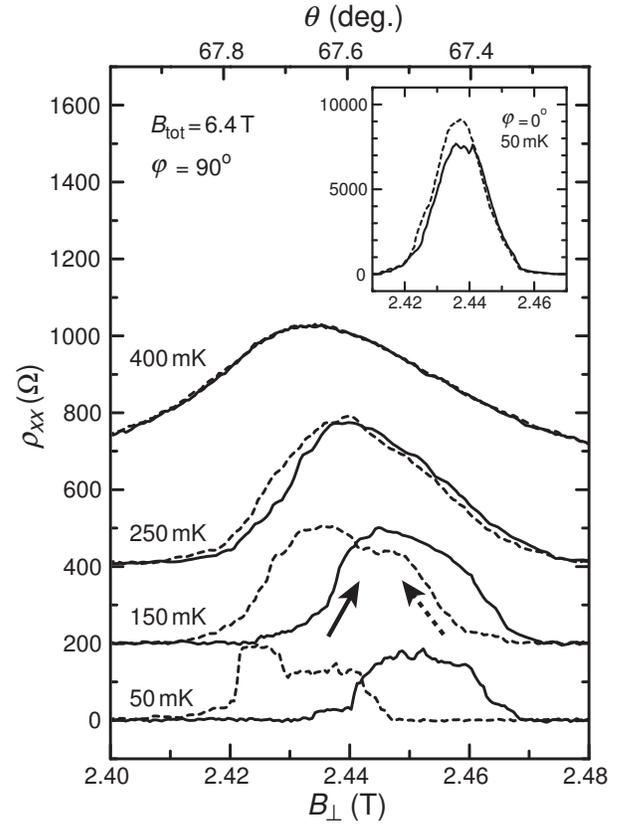}
\caption{
Hysteresis of the $\theta$-dependence of $\rho_{xx}$ plotted as a function of $B_\perp$ for $\varphi=90^\circ$.
Solid (dashed) lines represent the data obtained by increasing (decreasing) $B_\perp$ for different temperatures.
The curves are offset by $200~\Omega$ for clarity.
The inset represents the data for $\varphi=0^\circ$ obtained at $T=50$~mK.
}
\end{figure}
At 50~mK, the width of the hysteresis loop is very large and comparable to the width of the transition region with $\rho_{xx}>0$ \cite{HysteresisSdH}.
While a typical sweep rate $dB_\perp/dt$ was chosen to be $0.1~{\rm mT}/{\rm s}$, the hysteresis is highly reproducible and insensitive to the sweep rate.
Unlike the transition between fractional QH states in GaAs 2DESs \cite{Eom2000,Smet2002,Kraus2002},
the effect of nuclear spin is considered to be negligible due to poor hyperfine coupling.
Relaxation behavior of $\rho_{xx}$ was not observed when the sweep was stopped in the transition region.
Thus we consider that the hysteresis can be attributed only to Ising ferromagnetic domain formation.
The hysteresis becomes weaker as $T$ increases and invisible at 400~mK.
The disappearance of hysteresis with increasing $T$ was also observed for resistance spikes in the Shubnikov-de Haas oscillations of AlAs \cite{Poortere2000,Poortere2003} and (Cd, Mn)Te \cite{Jaroszynski2002} QH systems.
The temperature corresponding to the onset of hysteresis was considered to be equal to \cite{Jaroszynski2002} or lower than \cite{Poortere2003} the Curie temperature.
In these works, the amplitude of the resistance spike approaches zero as $T$ goes to zero.
On the other hand, the resistance peak remains even at 50~mK in the Si QH system.
The $T$-dependence of the peak height, which also depends on hysteresis, was found to be less than 15~\% for $\varphi=0^\circ$ from 50 to 300~mK.

A possible explanation for excess resistance at an Ising ferromagnetic transition is one-dimensional conduction along domain boundaries.
In Ref.~\cite{Jungwirth2001}, the Hartree-Fock quasiparticle energy was calculated to be reduced by nearly 50~\% at the center of the domain wall formed in AlAs 2DESs used in Ref.~\cite{Poortere2000}.
If the energy gap in the domain wall is very small in the present Si 2DES and the potential fluctuations overcome it, it seems possible that quasiparticles in the one-dimensional channels exist even at $T=0$.
The observed anisotropy of $\rho_{xx}$ corresponds to large longitudinal conductivity $\sigma_{yy}$ in the direction perpendicular to the in-plane magnetic field and implies a stripe-like domain structure oriented along this direction.

The resistance peak at the coincidence decreases rapidly as $T$ increases above 0.3~K.
We believe that this is due to the collapse of the Ising ferromagnetic domain structure.
In Fig.~3(a), $\rho_{xx}$ values for $\varphi=0^\circ$ after subtracting the baseline, which arises from the usual Shubnikov-de Haas oscillations, are plotted.
\begin{figure}[t!]
\includegraphics[width=8cm]{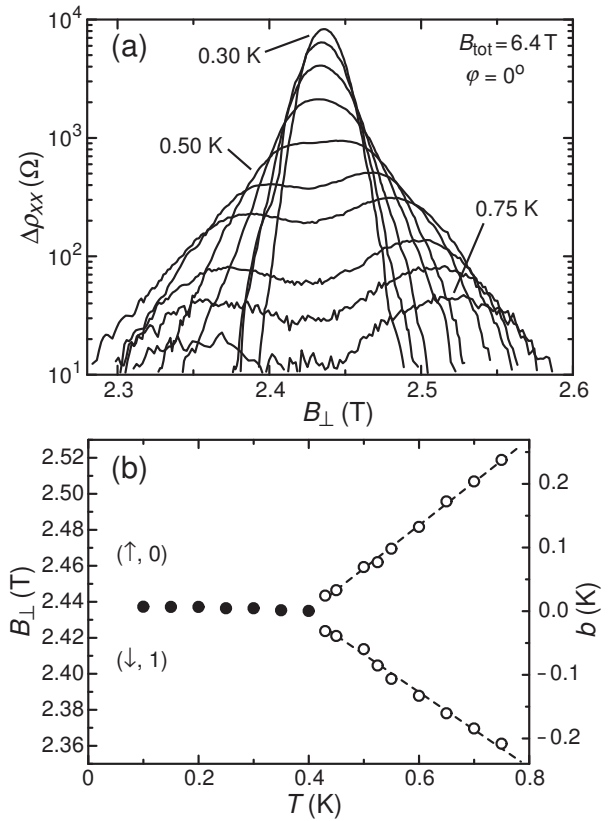}
\caption{
(a) Excess resistivity for $\varphi=0^\circ$ at different temperatures from 0.30~K to 0.75~K (from top to bottom) in steps of 0.05~K.
(b) Peak positions determined by a fit to a single Gaussian function (solid circles) or a double Gaussian function (open circles).
For low temperatures ($T<0.25$~K), where the hysteresis was observed, the mean values are plotted.
Dashed straight lines represent least-squares fits to the double peaks.
The right axis indicates the pseudospin Zeeman field $b=(\hbar e B_\perp/m^\ast-g^\ast \mu_B B_{\rm tot})/2$ (see text).
}
\end{figure}
A double peak structure is seen for $T \gtrsim 0.5$~K.
We do {\it not} consider that it is caused by the splitting of the valley degeneracy since this does not explain the strong positive $T$ dependence of the peak separation \cite{Valley}.
The observed double peak structure is well fitted with a couple of Gaussian functions for $T>0.40$~K.
The baseline subtraction procedure does not significantly change the peak positions since the baseline magnetoresistance is gradual and monotonic in the relevant range.
The obtained peak positions are plotted in Fig.~3(b) together with single peaks at lower temperatures.
Both peaks shift almost linearly with $T$.
The slope $dB_\perp/dT$ is 0.24~T/K for the higher $B_\perp$ peak and $-0.21$~T/K for the lower $B_\perp$ peak.

In order to discuss the pseudospin magnetization in the paramagnetic regime, we define the effective field as $b=(\hbar \omega_c-g^\ast \mu_B B_{\rm tot})/2$.
The pseudospin Zeeman energy can be written as $-b \sigma_z$ with $\sigma_z =+1$ for the LL$(\uparrow, 0)$ state and $\sigma_z =-1$ for the LL$(\downarrow, 1)$ state.
The scale of the right axis of Fig.~3(b) is chosen so that the single peak at $T=0.40$~K agrees with $b=0$, and an enhanced effective mass $m^\ast=0.24m_e$ \cite{Enhancedmass} is used.
We assume that $g^\ast$ remains almost constant in the narrow range of $B_\perp$.
In order to estimate the average pseudospin magnetization $\langle \sigma_z \rangle$ at the $\rho_{xx}$ peaks, we use the mean field equation
\begin{eqnarray}
\langle \sigma_z \rangle= \tanh \left( \frac{b+T_c \langle \sigma_z \rangle}{T} \right)
.
\end{eqnarray}
Here $T_c$ is the Curie temperature and the equation has nonzero solutions even at $b=0$ for $T<T_c$.
It can be transformed into
\begin{eqnarray}
b=T \tanh^{-1} \langle \sigma_z \rangle -T_c \langle \sigma_z \rangle
.
\end{eqnarray}
For a constant value of $\langle \sigma_z \rangle$, $b$ changes linearly with $T$.
Applying Eq.~(2) for the observed $T$ dependence of the peak positions, $\langle \sigma_z \rangle$ and $T_c$ are deduced to be 0.59 ($-0.53$) and 0.46~K (0.43~K) for the higher (lower) $B_\perp$ peak, respectively.
The results indicate that electron scattering in the {\it partially} pseudospin-polarized state is stronger than those in the unpolarized state and the fully polarized state.
It is suggested that thermal fluctuations of local pseudospin magnetization are related to electron scattering.

Finally we discuss the $I$-$V$ characteristics.
In order to avoid the heating of the substrate, the sample was immersed in liquid ${}^3$He.
In Fig.~4, the data obtained at $b=0$ and $\nu=4.0$ for $\varphi=0^\circ$ are shown.
\begin{figure}[t!]
\includegraphics[width=8cm]{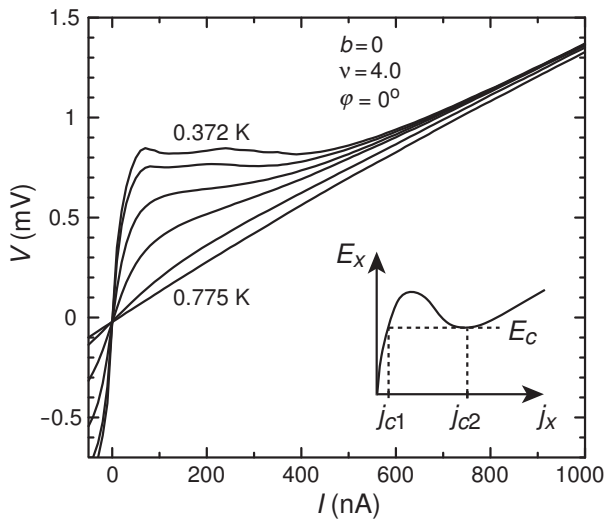}
\caption{
Current-voltage characteristics at $b=0$ and $\nu=4.0$ for $\varphi=0^\circ$.
The temperatures are (from top to bottom) 0.372, 0.433, 0.491, 0.554, 0.641, and 0.775~K.
The inset shows a possible relationship between the local electric field and current density.
}
\end{figure}
At 0.372~K, $V$ increases steeply with $I$ for $I < 60~{\rm nA}$, but it exhibits a nearly constant value over a broad current range $60~{\rm nA} < I < 450~{\rm nA}$. For $I > 450~{\rm nA}$, $V$ again increases with $I$ and the slope approaches that for higher temperatures.
The most striking feature is the voltage plateau in the intermediate region.
It can be understood in terms of a model based on an $N$-shaped relationship between the local electric field $E_x$ and current density $j_x$ as illustrated in the inset of Fig.~4.
It is expected that $\rho_{xx}=E_{x}/j_{x}$ decreases as $j_{x}$ increases due to the collapse of the Ising ferromagnetic domain structure induced by electron heating.
A rapid decrease may cause a negative differential resistivity $dE_{x}/dj_{x}<0$ followed by a local minimum $E_{x}=E_{c}$ at an upper critical current density $j_{c2}$.
The lower critical current density $j_{c1}$ is defined as the other solution of $E_{x}=E_{c}$.
When the average current density $I/w$ is in the range between $j_{c1}$ and $j_{c2}$, the uniform distribution of $j_x$ is not stable.
A separation into two regions with $j_x=j_{c1}$ and $j_x=j_{c2}$ should occur so as to minimize $E_x$ which is proportional to the Joule heating per unit length $E_x I$.
If this is the case, $V$ exhibits a constant value $E_{c} l$ in this range.

In summary, we have investigated transport properties of the Ising QH ferromagnet formed in a Si/SiGe heterostructure.
At the coincidence of the LLs, $\rho_{xx}$ exhibits a very strong anisotropy with respect to the in-plane magnetic field.
The resistance peak shows large hysteresis when the pseudospin effective field crosses zero at low temperatures.
Above $T_c$, we observed a double peak structure.
The linear $T$ dependence of the peak positions was discussed using the mean field theory.
The result indicates that electron scattering is strong when the pseudospin is partially polarized.
The $I$-$V$ characteristics, which exhibit a wide voltage plateau, were also studied.

We acknowledge useful discussions with D. Yoshioka and R. Masutomi, and thank A. Sekine for his technical assistance. 
This work was partly supported by Grant-in-Aid for Scientific Research (B) (No. 18340080), Grant-in-Aid for Scientific Research on Priority Area ``Physics of new quantum phases in superclean materials'' (No. 18043008), and Grant-in-Aid for JSPS Foundation (No. 1852232) from MEXT, Japan.


\begin{thebibliography}{30}
\bibitem{Crystallographic} We have also performed measurements on a Hall bar which was fabricated along the crystallographic direction perpendicular to that used in this work. No significant difference was found.
\bibitem{HysteresisSdH} No hysteresis was observed in ordinary Shubnikov-de Haas oscillations even at 50~mK.
\bibitem{Valley} The valley splitting $E_v$ in the present sample is estimated to be less than 0.2 K. At low magnetic fields where the many body effect is weak, a $\rho_{xx}$ minimum corresponding to the valley splitting does not appear while that for the Zeeman splitting can be observed even at a small perpendicular magnetic field of 0.29 T ($\nu = 34$). This indicates that the bare valley splitting is smaller than the energy gap $\Delta E=g^\ast \mu_B B-E_v$ between adjacent Landau levels having different spins.
\end{thebibliography}
\end{document}